\documentclass{article}
\usepackage{epsfig}
\usepackage{latexsym}
\usepackage{amsmath}
\usepackage{graphics}
\def\br{{\bf r}}
\def\bk{{\bf k}}

\begin{document}
\title{Depletion of a Bose-Einstein condensate by 
laser-induced dipole-dipole interactions}
\author{I.E. Mazets$^{1,3}$,
D.H.J. O'Dell$^2$,
G. Kurizki$^{3}$, N. Davidson$^{3}$ and W.P. Schleich$^{4}$ \\
$^1$Ioffe Physico-Technical Institute, St.Petersburg 194021, Russia \\
$^{2}$Department of Physics \& Astronomy,
University of Sussex, Brighton \\ BN1 9QH, United Kingdom \\
$^3$Weizmann Institute of Science,
76100 Rehovot, Israel \\
$^4$Abteilung f\"ur Quantenphysik, Universit\"at Ulm, Ulm D-89069, Germany.}
\date{}
\maketitle

\begin{abstract}
We study a gaseous atomic Bose-Einstein condensate with 
laser-induced dipole-dipole interactions  
using the Hartee-Fock-Bogoliubov theory within the
Popov approximation.
The dipolar interactions introduce long-range atom-atom
correlations which manifest themselves as
increased depletion
at momenta similar to that of the laser wavelength, as well
as a `roton' dip in the excitation spectrum.
Suprisingly, the roton dip and
the corresponding peak in the depletion are enhanced
by raising the temperature above absolute zero. 
\newline
PACS: 03.75.Fi, 05.30.Jp
\end{abstract}
%]

\section{Introduction}
One of the novel features of gaseous Bose-Einstein condensates (BECs),
when considered from a many-body physics perspective, is the 
ability to directly manipulate the interatomic interaction using external
electromagnetic fields \cite{burnett98}. In the BECs realized thus 
far the interatomic interactions can be described by a single parameter,
the s-wave scattering length, $a$, and are short-range in comparison to the
average interatomic distance.
The experiment by Inouye \textit{et al} \cite{inouye98} demonstrated
how the s-wave scattering length can be changed by magnetic fields
via a Feshbach resonance. Here we consider enhancing and
controlling the interatomic interactions using laser-induced
dipole-dipole forces. These interactions are intrinsically
long-range and so affect the gas in a way profoundly different
from a Feshbach resonance. In particular, dipole-dipole interactions,
both laser-induced \cite{odell03} and static \cite{santos03}, 
have been predicted to be capable of introducing a `roton'
dip into the Bogoliubov excitation spectrum of the BEC. This is 
behaviour reminiscent of that of the strongly correlated quantum 
liquid helium II, and can be understood
in terms of Feynman's relation which provides an upper bound 
for the spectrum $E(k)$ of helium II 
\begin{equation}
E(k) \le \frac{\hbar^{2} k^{2}}{2 m S(k)}
\label{eq:feynman}
\end{equation}
where $S(k)$, known as the static structure factor, is the Fourier 
transform of the pair distribution function. $S(k)$ is a measure of
2nd order correlation in the system. Feynman's formula interprets
the roton dip in helium II as being due to a peak in $S(k)$ and 
hence as being due to strong correlations, which are at 
values of $k$
corresponding to phonons having wavelengths on the order of the 
interatomic spacing (intuitively: the excitations at these wavelengths 
require less energy to excite in comparison
to neighbouring wavelengths which cost more compressional energy).

In a gaseous BEC at 
zero temperature the Bogoliubov theory of a 
weakly interacting degenerate Bose gas \cite{llstatphys2} is 
valid and shows that 2nd order
correlation arises predominantly from pairs of atoms scattering 
out of the condensate to form the so-called quantum depletion, 
or `above condensate' fraction.
In a homogenous system the atoms in these pairs are prefectly 
correlated, in the sense of perfect two mode-squeezing 
\cite{roberts,jesus,barnett}. However, in gaseous BECs
the quantum depletion, and hence 2nd order
correlation, is typically very small on account of 
their diluteness, as expressed through the gas parameter 
$na^{3} \ll 1$, where $n$ is the atomic density. 
Indeed, within the Bogoliubov theory the Feynman relation
(\ref{eq:feynman}) is an exact equality, yet in a measurement of the 
bulk excitation spectrum of a regular gaseous BEC interacting via s-wave 
scattering by Steinhauer \textit{et al} \cite{steinhauer02} 
no roton was seen.
Laser-induced dipole-dipole interactions, on the other hand, introduce 
\emph{controllable} (via laser intensity etc.) long-range correlations 
on the scale of the laser wavelength, and one 
might expect some significant depletion at the corresponding momentum.

In our previous investigations into laser-induced
dipole-dipole interactions in a BEC \cite{odell03,springerbook} (and 
references therein) we considered
the zero-temperature case, and the calculations were limited to
the basic Bogoliubov method \cite{llstatphys2} which does not
treat the depletion self-consistently.
Our purpose here is to use the so-called Popov version of the 
Hartree-Fock-Bogoliubov method
\cite{popov,griffin96}, which treats the depletion as a self-consistent
mean-field in order to include the effects of the the back-action 
of the depleted fraction upon the condensate.
This method was primarily developed to treat
the finite temperature case where depletion is important, 
but our hope here is to account for significant quantum depletion
due to enhanced interactions and also for thermal depletion due to non-zero
temperature. 
The most significant results we find here are that, perhaps contrary
to expectation, the depletion can increase the effects of the dipole-dipole
interactions, deepening the roton dip in the spectrum, so that
the roton is not diminished by finite-temperature effects but
can actually be enhanced.

\section{Laser induced dipole-dipole interactions}
%%%%%%%%%%%%%%%%%  F I G U R E 1 %%%%%%%%%%%%%%%%%%%%%%
%\begin{figure}[htbp]
%\begin{center}
%\centerline{\epsfig{figure=odellhfbfig1.eps,width=8cm}}
%\end{center}
%\vspace{-4ex} \caption{The laser beam and condensate geometry.}
%\label{fig:setup}
%\end{figure}
%%%%%%%%%%%%%%%%%%%%%%%%%%%%%%%%%%%%%%%%%%%%%%%%%%%%%%
The idea of exploiting dipole-dipole forces to modify the 
properties of atomic BECs first arose in the context of static
dipolar interactions.
A BEC of atoms with a large permanent magnetic moment
(e.g.\ chromium) ordered by an external magnetic field, or
equivalently, polarizable atoms in a static electric field, has
been considered by a number of authors
\cite{yi,goral,martikainen,santos2000,goral2002,goral2002b}.
Here, however, we are interested in fully retarded 
\emph{dynamic} dipole-dipole
interactions, such as those induced by an electromagnetic wave
(e.g. laser beam). The dynamic dipole-dipole interaction is
distinguished from the static case by a longer range (the
retardation gives it an attractive $r^{-1}$ component which can be
used in certain geometries to mimic gravity \cite{odell2000}) and
a huge enhancement of atomic polarizabilty close to a resonance.

\begin{figure}
\begin{center}
\centerline{\epsfig{figure=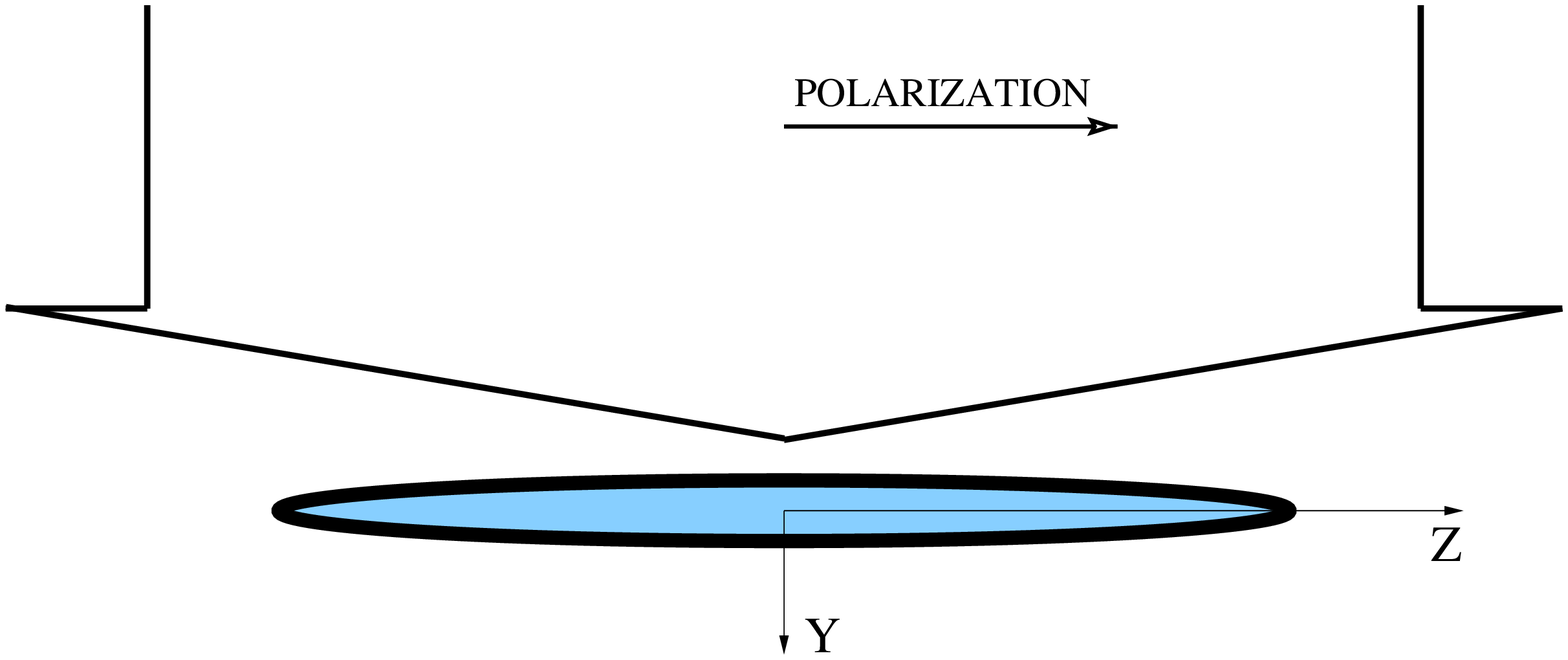,width=8cm}}
\end{center}
\vspace{-4ex} \caption{The laser beam and condensate geometry.}
\label{fig:setup}
\end{figure}

We consider a cigar-shaped BEC
tightly confined in the radial ({\it x,y}) plane, irradiated by a
plane-wave laser as in \cite{odell03}
(see Fig.\ \ref{fig:setup}). The laser
polarization is along the long z-axis of the condensate to
suppress collective (``superradiant'') Rayleigh scattering
\cite{inouye99} or coherent atomic recoil lasing
\cite{piovella2001} that are forbidden along the direction of
polarization. The tight confinement along the propagation
direction, together with the far off-resonance condition, enables
us to treat the electromagnetic field within the Born
approximation. The dipole-dipole potential induced by far off-resonance
electromagnetic radiation of intensity $I$, wave-vector ${\bf
k}_{\mathrm{L}}= k_{\mathrm{L}} \hat{{\bf y}}$ (along the y-axis),
and polarization $\hat{{\bf e}} = \hat{{\bf z}}$ (along the
z-axis) is \cite{thirunamachandran80}
\begin{equation}
U_{\mathrm{dd}}({\bf r}) = \frac{ 
I\alpha^{2}\left(\omega \right)}{4 \pi c \varepsilon_{0}^{2}}
\frac{1}{r^{3}}  \Big[ \big(1 - 3 \cos^2(\theta)\big) \big(
\cos \left(k_{\mathrm{L}} r \right)+  k_{\mathrm{L}}r \sin
 \left(k_{\mathrm{L}} r \right) \big) 
  -  \sin^2(\theta)  k_{\mathrm{L}}^{2}r^{2}
\cos  \left(k_{\mathrm{L}}r \right)\Big]
\cos \left(k_{\mathrm{L}} y
\right).   \label{eq:tpot}
\end{equation}
Here $\bf{r}$ is the interatomic axis, and 
$\theta$ is the angle between $\bf{r}$ and the
z-axis.  The far-zone ($ k_{\mathrm{L}}r \gg 1$) behavior of
(\ref{eq:tpot}) along the z-axis is proportional to $- \sin (
k_{\mathrm{L}}r) / ( k_{\mathrm{L}}r)^2$.  In terms of the
condensate density $n(\mathbf{r})= \vert \psi(\mathbf{r})
\vert^{2}$, the mean-field energy functional accounting for
atom-atom interactions is taken to be the sum $H_{\mathrm{dd}}
+H_{\mathrm{s}}$ \cite{goral,yi,odell2000}, where $H_{\mathrm{dd}}
= (1/2) \int n(\br)
 \, U_{\mathrm{dd}}(\br-\br')\,  n(\br')
 \, d^{3}r \, d^{3}r'$, and $H_{\mathrm{s}}  =  (1/2) (4
\pi a \hbar^{2}/m) \int n^{2}(\br) d^{3}r$ is due to short-range
($r^{-6}$) van der Waals interactions, which are described, as is
usual, by a delta function pseudo-potential $U_{\mathrm{s}}(\br)=(4
\pi a \hbar^{2}/m) \delta(\br)$.
In momentum space $H_{\mathrm{dd}}$ takes the form,
$H_{\mathrm{dd}}= (1/2)(2 \pi)^{-3}$ $ \int
\widetilde{U}_{\mathrm{dd}}({\mathbf k}) \, \tilde{n}(\bk) \,
\tilde{n}(-\bk) \, d^{3}k$, where 
$\widetilde{U}_{\mathrm{dd}}({\mathbf k})= \int  \, \exp
[-{\mathrm i} {\mathbf k} \cdot {\mathbf r}] \,
U_{\mathrm{dd}}({\mathbf r}) \, d^{3}r$ is
the Fourier transform of the
dipole-dipole potential (\ref{eq:tpot}),
the explicit form of which, for the laser propagation and
polarization as in Fig.~\ref{fig:setup}, we have given in Eq.\ (4) of
\cite{odell03}.

\begin{figure}
\begin{center}
\centerline{\epsfig{file=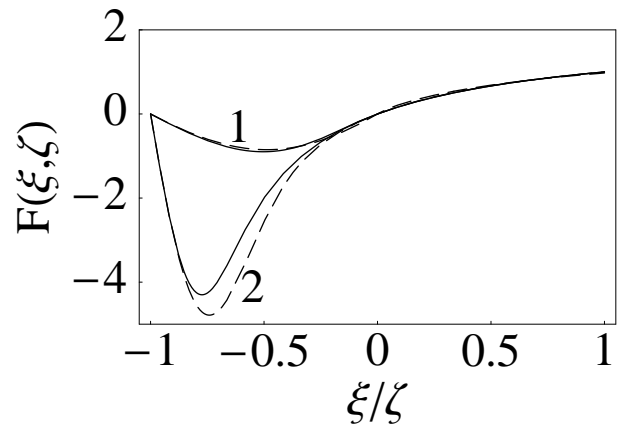,width=8cm}}
\end{center}
\caption{
Comparison of the exact (dashed line) and approximate 
(solid line) expressions for $F(\xi , \zeta )$ for (1) $\zeta =2.9$, 
the solid and dashed lines are practically indistinguishable, (2) 
$\zeta  =32$, the difference between the two lines is small but visible.} 
\label{fig:exact+appr}
\end{figure}

When the radial trapping is sufficiently tight so that the BEC
is in its radial ground state we may adopt a cylindrically
symmetric gaussian ansatz for the density profile
whose width, $w_{r}$, is 
\begin{equation}
n^{\mathrm{3D}}(\br) \equiv N (\pi w_{r}^{2}
)^{-1} n(z) \exp \left[-(x^{2}+y^{2})/w_{r}^{2} \right] ,
\end{equation}
where N is the total number of atoms and $n(z)$, the 1D axial density,
is normalized
to 1, but is unspecified for the time being. Denoting by
$\tilde{n}^{\mathrm{3D}}(\bk)$ the Fourier transform of 
$n^{\mathrm{3D}}(\br)$, we have
$\tilde{n}^{\mathrm{3D}}(\bk)= N \widetilde{n}(k_{z})
 \exp \left[-w_{r}^{2}(k_{x}^{2}+k_{y}^{2})/4
\right]$, in which $\widetilde{n}(k_{z})$ is the Fourier
transform of the axial density $n(z)$. This ansatz allows the
principal value of the radial integration in $H_{\mathrm{dd}}$ to
be evaluated analytically so that the dipole-dipole energy reduces
to a one dimensional functional along the axial ($\hat{z}$) direction
\begin{equation}
H_{\mathrm{dd}} = (N^{2}/2) \int n(z)
 n(z') U^{z}_{\mathrm{dd}}(z-z') \, dz \, dz'
 =   (N^{2}/4 \pi) \int \widetilde{n}(k_{z})
\widetilde{n}(-k_{z}) \widetilde{U^{z}_{\mathrm{dd}}}(k_{z})
\, d k_{z} .
\end{equation} 
Defining the variables 
\begin{equation}
\zeta =k_{\mathrm{L}}^2w_r^2/2 \quad , \quad 
\xi =(k_z^2-k_{\mathrm{L}}^2)w_r^2/2 ,
\end{equation} 
the one-dimensional (1D) axial 
potential that appears in $H_{\mathrm{dd}}$ is
\begin{equation}
\widetilde{U^{z}_{\mathrm{dd}}}(k_{z})  =  \frac{ I \,
\alpha^{2} k_{\mathrm{L}}^2}{4 \pi \epsilon_{0}^{2} \; c} \,
Q[\xi(w_r,k_{z}) ,\zeta(w_r)] 
,  \label{ueff1}
\end{equation}
where $
Q[\xi(w_r,k_{z}) ,\zeta(w_r)]  
=  \frac 1{2 \zeta}\left[  -\frac 23 +F(\xi, \zeta)\right],
$ and 
\begin{equation}
F(\xi, \zeta)=2\xi \,\exp ({\xi - \zeta })\sum _{j=0}^\infty
\frac {\zeta ^j}{j!}\, \Re  \{E_{j+1}(\xi )\}.
\label{ueff2}
\end{equation}
$ \Re \{ E_{j}[z] \} $ is the real part of the generalized
exponential integral \cite{a+s}. 
However, the sum in the expression (\ref{ueff1}) is rather unwieldy to use 
in calculations, so instead we will substitute it by
the
following asymptotic representation for the function $F$:
\begin{equation}
F(\xi, \zeta )\approx 
\left \{
\begin{array}{lr}
\frac {2\xi }{\zeta +\xi } \left \{ 1 -
\exp \left[ (\zeta +\xi )\left( 1-\sqrt{ -\frac \zeta \xi } \right)
\right] \right \} ,& \xi <0 \\
\frac {2\xi }{\zeta +\xi }, & \xi \ge 0
\end{array}
\right. .
\end{equation}
This representation is very good for $\xi >0$ and gives the correct
asymptotics for $\xi \rightarrow 0$ ($k_z\rightarrow k_L$) and
$\xi \rightarrow -\zeta $ ($k_z\rightarrow 0$). For $-\zeta <\xi <0$
the agreement is reasonably good (see Figure \ref{fig:exact+appr}).

%%%%%%%%%%%%%%%%%%%%%%%% FIGURE   2 %%%%%%%%%%%%%%%%%%%%%%%%%%%%%%%%%%%%%%%%
%\begin{figure}
%\begin{center}
%\centerline{\psfig{file=odellhfbfig2.eps,width=8.cm}}
%\end{center}
%\caption{
%Comparison of the exact (dashed line) and approximate 
%(solid line) expressions for $F(\xi , \zeta )$ for (1) $\zeta =2.9$, 
%the solid and dashed lines are practically indistinguishable, (2) 
%$\zeta  =32$, the difference between the two lines is small but visible.} 
%\label{fig:exact+appr}
%\end{figure}
%%%%%%%%%%%%%%%%%%%%%%%%FIGURE%%%%%%%%%%%%%%%%%%%%%%%%%%%%%%%%%%%%%%%%%

Using the above expression, the Fourier Transform of the total (s-wave
plus dipole-dipole) 1D reduced interatomic potential is
\begin{equation}
\widetilde{U^{z}_{\mathrm{tot}}}(k_{z}) =4 E_{\mathrm{R}}
 a \left((k_{\mathrm{L}} w_{r})^{-2} +
\mathcal{I}Q(w_{r}, k_{z}) \right) \label{eq:tot1dpot}
\end{equation}
where $E_{\mathrm{R}}= \hbar^{2} k_{\mathrm{L}}^{2}/ 2m$ is the
photon recoil energy of an atom and $\mathcal{I}$ is the
dimensionless laser `intensity' parameter
\begin{equation}
{\mathcal I}= \frac{ I \, \alpha^{2}(\omega) m}{8 \pi
\varepsilon_{0}^{2} c \hbar^{2} a}. \label{eq:itildedefn}
\end{equation}
It is emphasized that the radial degree of freedom is contained in
(\ref{eq:tot1dpot}) via the radius $w_{r}$. Note that the Fourier 
Transform of the total potential is an even function: 
$\widetilde{U^z_{\mathrm{tot}}}(k_z)=\widetilde{U^z_{\mathrm{tot}}}
(-k_z)$.

\section{Hartree-Fock-Bogoliubov-Popov method} 
\label{sec:HFBP}

The Hartree-Fock-Bogoliubov (HFB) method is concisely described in 
\cite{griffin96}. A reasonable approach within the HFB method is the 
Popov approximation \cite{popov}, which neglects the contribution of 
the above-condensate fraction to the anomalous correlation function. 
The advantage of the Popov approximation is that it gives a gapless 
spectrum of elementary excitations of a degenerate bosonic system 
in a wide temperature range (up to the critical temperature). 
In the present paper we apply the Popov approximation to a quasi-1D 
BEC with the laser-induced dipole-dipole interactions 
at finite temperature. Of course, for an infinitely long 1D gas there 
is no condensate, even if this gas is non-ideal. This is because the 
depletion diverges at infinitely long wavelength. However, taking into 
account the finite length $\ell $ of the sample, we set an effective 
cutoff at a wavelength $\sim \ell $ and thus remove the divergence. 
Hence, a macroscopic population of the ground state becomes possible (in 
the case of an interacting gas). 

The 1D ansatz for the atomic field operator is $\hat \Psi (x,y,z,t)=
\hat \psi (z,t) \exp [-(x^2+y^2)/(2w_r^2)]/(\sqrt{\pi }w_r)$. Then the 
1D Hamiltonian reads as 
\begin{equation} 
\hat H= \int dz\,  \hat \psi ^\dag (z)\left( -\frac {\hbar ^2}{2m} 
\frac {\partial ^2}{ \partial z^2}\right) \hat \psi (z)+
\frac 12 \int dz\int dz^\prime \, \hat \psi ^\dag (z^\prime ) 
\hat \psi ^\dag (z) U^z_{\mathrm{tot}}(z-z^\prime )\hat \psi (z) 
\hat \psi (z^\prime  ), 
\label{H1D}
\end{equation} 
where $U^z_{\mathrm{tot}}(z)$ is the inverse Fourier Transform of 
$\widetilde{U^z_{\mathrm{tot}}}(k_z)$ defined by 
Eq.~(\ref{eq:tot1dpot}). Decomposing the atomic field operator in 
plane waves $\hat \psi (z)=\sum _k \ell ^{-1/2}\exp (\mathrm{i}kz) 
\hat a_k$ we 
rewrite Eq.~(\ref{H1D}) in the momentum   space: 
\begin{equation} 
\hat H=\sum _k \frac {\hbar ^2k^2}{2m}\hat a^\dag _k \hat a_k +
\frac 1{2\ell }\sum _{k,k^\prime ,q}\widetilde{U^z_{\mathrm{tot}}}(q)
\hat a^\dag _{k+q}\hat a^\dag _{k^\prime -q} \hat a_{k^\prime} \hat a_k.
\label{H1Da}
\end{equation}
The Hamiltonian of Eq.\ (\ref{H1Da}) applies to a system with a fixed 
number of particles. For simplicity we will work within the 
grand canonical ensemble and use the chemical potential, $\mu$,  
to fix the mean particle number $N$. 
Then the Heisenberg equation of motion for the atomic annihilation 
operators $\hat{a}_{k}$,  is given by the commutator of $\hat{a}_{k}$ 
with $\Hat{H}-\mu \sum_{k}\hat{a}^{\dag}_{k} \hat{a}_{k}$, and reads
\begin{equation}
\mathrm{i} \hbar \frac \partial {\partial t}\hat a_k=\left( 
\frac {\hbar ^2k^2}{2m}-\mu \right) 
\hat a_k +\frac 1\ell  
\sum _{k^\prime ,q}\widetilde{U^z_{\mathrm{tot}}}(q)
\hat a^\dag _{k^\prime +q}\hat a _{k^\prime  }\hat a_{k+q} .
\label{eq:heisa1}
\end{equation} 
Since the state with zero momentum is macroscopically populated, we 
invoke broken symmetry arguments \cite{llstatphys2} and replace 
$\hat a_0$ by a c-number 
$\sqrt{N_c}\exp (\mathrm{i}\varphi )$, where $N_c$ is the number of atoms 
in the condensate. The phase $\varphi $ is not an observable quantity, 
and its choice is arbitrary, thus we set $\varphi =0$. The 
linear (1D) condensate 
density is defined as $n_c=N_c/\ell $. Atoms with $k\neq 0$ comprise the 
above-condensate fraction (i.e., depletion) with the 
linear density 
\begin{equation}
n_a =\frac 1\ell \sum _{k\neq 0}\left \langle  \hat a_k^{\dag}\hat a_k 
\right \rangle ,
\end{equation}
which, together with $n_c$, yields the total linear density  
$$n=n_c+n_a\equiv N/\ell .$$ 

Finally, in the Popov approximation the chemical potential is given 
by the expression 
\begin{equation} 
\mu = \widetilde{U^z_{\mathrm{tot}}} (0)(n_c+  
n_a)+\widetilde{W}(0),  
\label{eq:mu}
\end{equation} 
where 
\begin{equation} 
\widetilde{W}(k)=\frac 1\ell \sum _{k^\prime \neq -k}
\widetilde{U^z_{\mathrm{tot}}}(k^\prime )\left \langle 
\hat a ^\dag_{k^\prime +k}\hat a _{k^\prime +k}\right \rangle 
.\label{eq:W}
\end{equation} 
For the modes with $k\neq 0$ we obtain, from Eq.~(\ref{eq:heisa1})  
\begin{equation}
\mathrm{i}\hbar \frac \partial {\partial t} \hat a_k = \left( \frac {\hbar ^2
k^2}{2m}-\mu \right) \hat a_k +\left[ \widetilde{U^z_{\mathrm{tot}}}
(0)+\widetilde{U^z_{\mathrm{tot}}}(k)\right] n_c\hat a_k +
 \left[ \widetilde{U^z_{\mathrm{tot}}}(0)n_a+\widetilde{W}(k) 
\right] \hat a_k +\widetilde{U^z_{\mathrm{tot}}}(k)n_c
\hat a^\dag _{-k}.
\label{eq:heisa2}
\end{equation}
When deriving Eq.~(\ref{eq:heisa2}), we take into account that 
$\left \langle \hat a^\dag   _k\hat a _{k^\prime }\right \rangle =
\left \langle \hat a^\dag   _k\hat a _{k}\right \rangle 
\delta _{k^\prime \,k}$. Also note that the anomalous correlation 
functions for the above-condensate operators, such as 
$\left \langle \hat a_k\hat a _{k^\prime }\right \rangle $, are 
neglected in Eqs.~(\ref{eq:mu},~\ref{eq:heisa2}), as required by the 
Popov approximation. 

We now implement the standard Bogoliubov transformation 
\cite{llstatphys2} 
$$
\hat a_k=u_k \exp (-\mathrm{i}\omega _kt)\hat b_k
-v_k\exp (\mathrm{i}\omega _kt)
\hat b^\dag _{-k},
$$
where the new (quasiparticle) operators $\hat b^\dag _k,\, \hat b_k$ 
obey the usual bosonic commutation relations. The solution of the 
generalized eigenvalue problem gives the dispersion relation 
\begin{equation}
\hbar \omega _k=\sqrt{T(k)\left[ T(k)+2\widetilde{U^z_{\mathrm{tot}}}
(k)n_c\right] }, 
\label{eq:disp}
\end{equation}
where 
\begin{equation}
T(k)=\frac {\hbar ^2k^2}{2m}+\widetilde{W}(k)-\widetilde{W}(0).
\label{eq:defT}
\end{equation}
The transformation coefficients are 
\begin{equation}
u_{k}= \sqrt{\frac{T(k)
+\widetilde{U^{z}_{\mathrm{tot}}}(k) \ n_{a}}{2 \hbar \omega_{k}}
+ \frac{1}{2}} \quad , \quad v_{k}= \sqrt{\frac{T(k)
+\widetilde{U^{z}_{\mathrm{tot}}}(k) \ n_{a}}{2 \hbar \omega_{k}}
- \frac{1}{2}} .
\label{eq:u,v}
\end{equation}
Note that if $\widetilde{U^z_{\mathrm{tot}}}(k)$ were $k$-independent 
(as it is in the case of absence of the external laser radiation, 
${\mathcal I}=0$) then $T(k)$ would coincide with the free atom 
kinetic energy, and the well-known dispersion relation for a BEC 
with short-range interactions in the Popov approximation 
\cite{griffin96,popov} would hold. 
At a finite temperature $\Theta $ (measured in energy units) we have 
\cite{griffin96} 
\begin{equation}
\left \langle \hat a^\dag _k\hat a_k\right \rangle =
u_k^2 [\exp (\hbar \omega _k/\Theta ) -1]^{-1}+v_k^2 . \label{eq:dep23}
\end{equation}

The Hartree-Fock-Bogoliubov-Popov equations should be solved 
self-consistently to obtain the condensate 
and above-condensate occupation numbers, under the constraint
that the total density is fixed, $n=n_{c}+n_{a}$. 
This involves an iterative procedure \cite{griffin96}:  \\
1) As a starting point we choose the occupation  
numbers given by the $\{u_{k},v_{k}\}$ coefficients 
found from the basic Bogoliubov scheme \cite{llstatphys2} 
(which does not treat the depletion self-consistently).\\
2) The occupation numbers are used to calculate
$\widetilde{W}(k)$ [Eq.\ (\ref{eq:W})] and hence the dispersion
relation $\hbar \omega_k$ [Eq.\ (\ref{eq:disp})].\\
3) The occupation numbers are summed to give the new total
density $n^{\prime}$, which is used to renormalize the 
occupation numbers of the condensate and above condensate modes, 
in order to keep the total density fixed. \\
4) The dispersion relation, incorporating the new occupation numbers, 
is then used to recalculate the $\{u_{k},
v_{k} \}$ coefficients [Eq.\ (\ref{eq:u,v})].\\
5) Repeat above steps 2--4 until convergence is achieved.

\section{Results  and Discussion }
\begin{figure}
\begin{center}
\centerline{\epsfig{figure=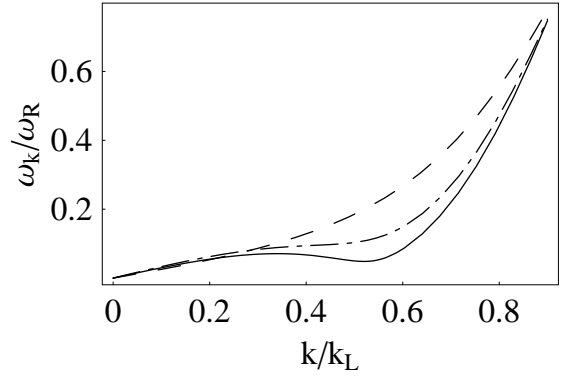,width=8cm}}
\end{center}
\vspace{-4ex} \caption{The spectrum of the elementary excitations
of a BEC with laser-induced dipole-dipole interactions, for the set 
of parameters of case (1), see text. The solid line corresponds to
the exact numerical solution, the dashed 
and dot-dashed lines correspond to 
the Bogoliubov spectrum for the variants (a) and (b), 
respectively. The frequency unit  
is the recoil frequency $\omega _{\mathrm{R}}=
E_{\mathrm{R}}/\hbar$. }
\label{fig:spec1}
\end{figure}

\begin{figure}
\begin{center}
\centerline{\epsfig{figure=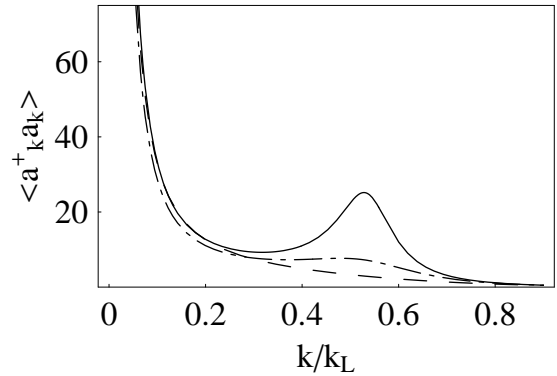,width=8cm}}
\end{center}
\vspace{-4ex} \caption{The depletion (above condensate fraction)
of a BEC with laser-induced dipole-dipole interactions, for the set 
of parameters of case (1), see text. The solid line corresponds to
the exact numerical solution, the dashed and dot-dashed lines correspond 
to the depletion calculated for the variants (a) and (b), respectively. 
Note the correspondence between the position of the peak in the mean
occupation number and that of the `roton' dip in Fig.\ \ref{fig:spec1}
(see Ref.\ \cite{odell03}).
}
\label{fig:dep1}
\end{figure}

\begin{figure}
\begin{center}
\centerline{\epsfig{figure=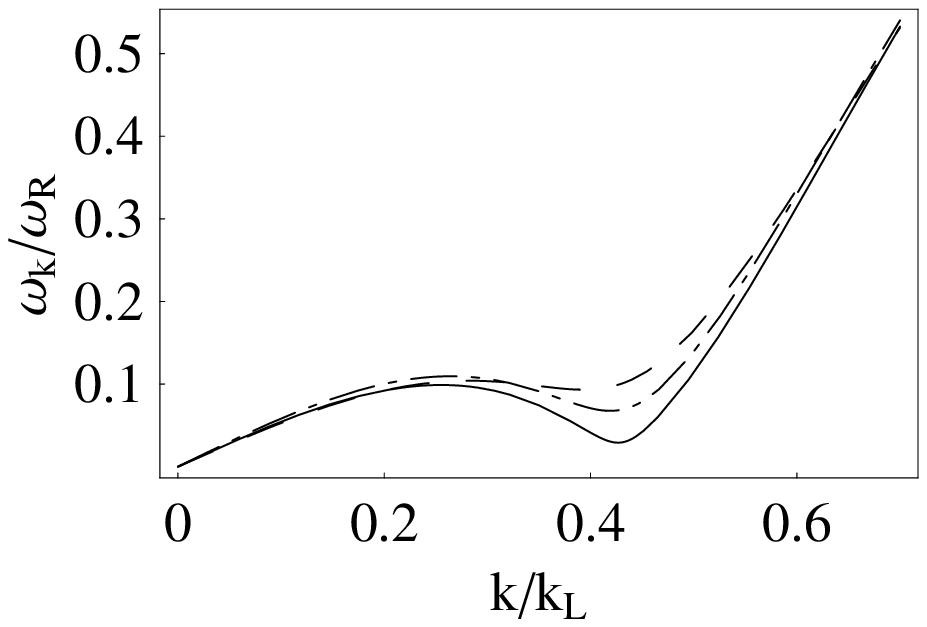,width=8cm}}
\end{center}
\vspace{-4ex} \caption{The same as in Fig.~\ref{fig:spec1} for the set 
of parameters of case (2), see text.}
\label{fig:spec2}
\end{figure}

\begin{figure}
\begin{center}
\centerline{\epsfig{figure=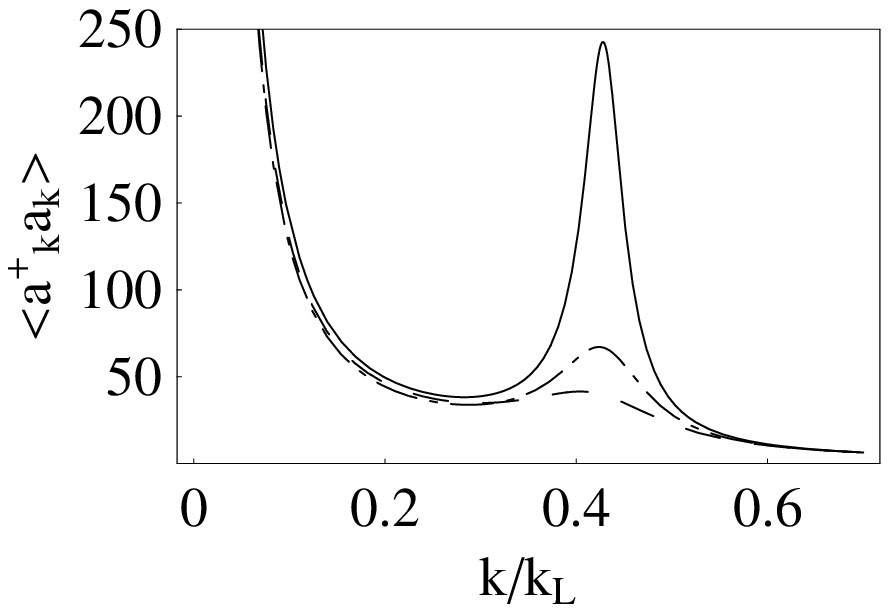,width=8cm}}
\end{center}
\vspace{-4ex} \caption{The same as in Fig.~\ref{fig:dep1} for the set 
of parameters of case (2), see text.}
\label{fig:dep2}
\end{figure}

We have solved the HFB equations with the Popov approximation, as described 
above for 
the basic setup shown in Figure \ref{fig:setup}. In Figures 
\ref{fig:spec1}--\ref{fig:dep2} we present the results for two sets 
of parameters: \\
\textit{Case 1}: $k_Lw_r=2.4$, $na=0.66$, $\ell /w_r=74$, $\Theta /
E_{\mathrm{R}}=0.68$, ${\mathcal I}=1.12$; \\
\textit{Case 2}: $k_Lw_r=8.0$, $na=4.7$, $\ell /w_r=28$, $\Theta /
E_{\mathrm{R}}=2.3$, ${\mathcal I}=0.28$. \\
For example, for $^{87}$Rb these sets of parameters mean explicitly: \\
\textit{Case 1}: The condensate radius $w_{r}$ is set at 
$0.3~\mu $m, the 3D density at the trap centre 
is equal to $4\cdot 10^{14}$ cm$^{-3}$, $\ell =22~\mu $m, 
$\Theta = 120$ nK.  The laser intensity can be found from the relation $I/
\Delta^{2}= 0.26~\mathrm{W/(cm}\cdot \mathrm{GHz)}^2$, where $\Delta $ 
is the laser detuning from resonance.  \\
\textit{Case 2}: $w_r=1.0~\mu $m, the peak 3D density is the same as 
in case 1, $\ell = 28~\mu $m, $\Theta =410$ nK,  $I/
\Delta^{2}= 0.064~\mathrm{W/(cm}\cdot \mathrm{GHz)}^2$.

Case 1 is representative of a BEC at the crossover between the ideal-gas 
and Thomas-Fermi limits with respect to the radial motion, while 
in case 2 the radial motion satisfies the conditions of the 
Thomas-Fermi limit. The parameters in cases 1 and 2 are chosen so that
a significant `roton' dip appears only in the more sophisticated 
Hartree-Fock-Bogoliubov-Popov method, but not in the simple Bogoliubov
approach.

In each of Figures \ref{fig:spec1}--\ref{fig:dep2} we have plotted 
three curves. The solid line gives the results of the 
Hartee-Fock-Bogoliubov-Popov calculation described in section 
\ref{sec:HFBP}. At convergence we found that in case 1 55~\% of atoms 
remain in the condensate and in case 2 the condensate fraction is 82~\%. 
The two other curves in the figures give the results obtained within the 
simple Bogoliubov approach and differ in the density that appears in 
the Bogoliubov dispersion relation. The variant (a) uses as the 
density in the Bogoliubov dispersion relation the  
condensate density $n_c$ obtained in the numerical calculations. The 
variant (b) uses in the same context as the density the total density 
of the system, $n=n_c+n_a$. Similarly, we calculate the depletion 
within the simple Bogoliubov approach for the variants (a) and (b). 
The position of the dip 
centre remains unchanged with temperature.

Figures \ref{fig:spec1} and \ref{fig:spec2} show that when the depletion
is taken into account self-consistently, the `roton' dip in the
excitation spectrum can actually become deeper. 
 This means that the 
`roton' dip persists at finite temperatures (of course, below the 
condensation temperature). Thus the sample heating due 
to spontaneous scattering of laser photons need not prevent observation of 
the `roton' dip in the spectrum of the elementary excitations. 
Contrary to what one might expect, finite temperature effects not only 
do not wash out or diminish these intermode correlations, but rather serve 
to enhance them. This is surprising, considering the noisy character
of finite-temperature density fluctuations. This might in part be
because the dipole-dipole potential has a momentum dependence
such that it is more effective at finite interatomic momenta
(see Fig. \ref{fig:exact+appr}). 

One may ask: what happens if the laser intensity is increased
so that the roton dip passes through the zero-energy axis, 
so that we obtain imaginary excitation frequencies? 
Two scenarios are then possible.  
The first one is that this would create 
a dynamical instability similar 
to that found in a BEC of atoms with negative scattering 
length \cite{aneg}. The instability would grow exponentially,  
finally leaving us with a totally depleted BEC. 
Another scenario is the following: when the laser 
intensity reaches the value  at which
the roton minimum touches the zero-energy axis we may  
access a new ground state that is  
periodically modulated along the $z$-axis: a supersolid.
An analogous behaviour had been predicted for superfluid liquid 
helium passing through a pipe \cite{pitaevskii84}.  
The formation of a supersolid state in a BEC with laser-induced
dipole-dipole interactions 
has recently been studied  \cite{sprsld}
by numerically integrating the 
time-dependent generalized Gross-Pitaevskii equation describing  
a cigar-shaped BEC irradiated by a circularly polarized laser
(in contrast to the linear case considered in the present paper),  
whose wave vector is along the major axis, $z$, of the cigar. 
This new supersolid state has to be stable against 
small perturbations. One can 
choose between these two scenarios by numerical  
investigation of the BEC dynamics in the case of 
gradually-increasing laser intensity, based on  
a time-dependent generalization of the Popov approximation. Such  
an investigation is beyond the scope of the present paper. 

Finally, we would like to briefly mention experimentally detectable
signatures of the 
roton minimum. Whilst the most convincing method would certainly be to map
out the dispersion formula using Bragg spectroscopy, as in the 
measurement by Steinhauer \textit{et al} \cite{steinhauer02}, Bragg spectroscopy is 
technically difficult and time
consuming. A simpler method would be to perform
a time-of-flight measurement after releasing the BEC from the trap in order
to detect the enhanced number of atoms having momenta around the roton minimum.

To summarize, we have presented a method allowing self-consistent 
calculation of the effects of laser-induced dipole-dipole interactions 
upon a BEC in the presence of significant condensate depletion. This
is especially relevant to the experimental situation where the temperature
cannot be significantly lower than the chemical potential.
This may help us cope with the adverse effects of the Rayleigh scattering
of laser light \cite{odell03}, which tend to heat up the sample.

\section{Acknowledgements}

We are grateful to the 
Engineering and Physical Sciences Research Council UK (EPSRC), the
German-Israeli Foundation (GIF), and the EU QUACS and CQG networks
for funding. I.E.M. also thanks the Russian funding sources 
(RFBR 02--02--17686, UR.01.01.040, E02--3.2--287).

\newpage

%%%%%%%%%%%%%%%%%  F I G U R E %%%%%%%%%%%%%%%%%%%%%%
%%%%%%%%%%%%%%%%%%%%%%%%%%%%%%%%%%%%%%%%%%%%%%%%%%%%%%

%%%%%%%%%%%%%%%%%%%%%%%% FIGURE 2 %%%%%%%%%%%%%%%%%%%%%%%%%%%%%%%%%%%%%
%%%%%%%%%%%%%%%%%%%%%%%%FIGURE%%%%%%%%%%%%%%%%%%%%%%%%%%%%%%%%%%%%%%%%%

%%%%%%%%%%%%%%%%%  F I G U R E  3  %%%%%%%%%%%%%%%%%%%%%%
%%%%%%%%%%%%%%%%%%%%%%%%%%%%%%%%%%%%%%%%%%%%%%%%%%%%%%

%%%%%%%%%%%%%%%%%  F I G U R E  4  %%%%%%%%%%%%%%%%%%%%%%
%%%%%%%%%%%%%%%%%%%%%%%%%%%%%%%%%%%%%%%%%%%%%%%%%%%%%%

%%%%%%%%%%%%%%%%%  F I G U R E  5 %%%%%%%%%%%%%%%%%%%%%%
%%%%%%%%%%%%%%%%%%%%%%%%%%%%%%%%%%%%%%%%%%%%%%%%%%%%%%

%%%%%%%%%%%%%%%%%  F I G U R E  6  %%%%%%%%%%%%%%%%%%%%%%
%%%%%%%%%%%%%%%%%%%%%%%%%%%%%%%%%%%%%%%%%%%%%%%%%%%%%%


\newpage


\begin{thebibliography}{10}
%\setlength{\itemsep}{0cm}

\bibitem{burnett98}
{K. Burnett, Nature \textbf{392}, 125
(1998).}

\bibitem{inouye98}
{S. Inouye, M.R. Andrews, J. Stenger, H.J. Miesner, D.M.
Stamper-Kurn and W. Ketterle, Nature \textbf{392}, 151 (1998).}

\bibitem{odell03}
{D.H.J. O'Dell, S. Giovanazzi, and G. Kurizki, Phys. Rev. Lett.
\textbf{90}, 110402 (2003).}

\bibitem{santos03}
{L. Santos, G.V. Shlyapnikov, and M. Lewenstein,
Phys. Rev. Lett. \textbf{90}, 250403 (2003).}

\bibitem{llstatphys2}
{E.M. Lifshitz and L.P. Pitaevskii, \textit{Statistical Physics
Part 2} (Butterworth-Heinemann, Oxford, 1998).}

\bibitem{roberts}
{T. Gasenzer, D.C. Roberts,  and K. Burnett, Phys. Rev. A
\textbf{65}, 021605 (2002).}

\bibitem{jesus}
{J. Rogel-Salazar, G.H.C. New, S. Choi and K. Burnett, Phys. Rev.
A \textbf{65}, 023601 (2002).}

\bibitem{barnett}
{S.M. Barnett and P.M. Radmore, \textit{Methods in Theoretical
Quantum Optics} (Clarendon press, Oxford, 1997).}

\bibitem{steinhauer02}
{J. Steinhauer, R. Ozeri, N. Katz and N. Davidson, Phys. Rev.
Lett. \textbf{88}, 120407 (2002).}

\bibitem{springerbook}
{G. Kurizki, S. Giovanazzi, D.H.J. O'Dell, and A.I. Artemiev in
\textit{Dynamics and Thermodynamics of Systems with Long-Range 
Interactions}, edited by T. Dauxois \textit{et al},
Lecture Notes in Physics
\textbf{602}, p382 (Springer, Berlin, 2002).}

\bibitem{popov} {V.N.~Popov, {\it Functional Integrals and Collective 
Modes} (Cambridge University Press, NY, 1987).} 

\bibitem{griffin96}
{A. Griffin, Phys. Rev. B \textbf{53}, 9341 (1996).}

\bibitem{yi}
{S. Yi and L. You, Phys. Rev. A \textbf{61}, 041604 (2000);
\textit{ibid.} \textbf{63}, 053607 (2001).}

\bibitem{goral}
{K. G{\'{o}}ral, K. Rz{\c{a}}{\.{z}}ewski, and T. Pfau, Phys. Rev.
A {\bf 61}, 051601 (2000).}

\bibitem{martikainen}
{J.-P. Martikainen, Matt Mackie, and K.-A. Suominen, Phys. Rev. A
\textbf{64}, 037601 (2001).}

\bibitem{santos2000}
{L. Santos, G.V. Shlyapnikov, P. Zoller, and M. Lewenstein, Phys.
Rev. Lett. \textbf{85}, 1791 (2000).}

\bibitem{goral2002}
{K. G{\'{o}}ral, L. Santos, and M. Lewenstein, Phys. Rev. Lett.
\textbf{88}, 170406 (2002).}

\bibitem{goral2002b}
{K. G{\'{o}}ral and L. Santos, Phys. Rev. A \textbf{66}, 023613
(2002).}

\bibitem{odell2000}
{D. O'Dell, S. Giovanazzi, G. Kurizki and V.M. Akulin, Phys. Rev.
Lett. {\bf 84}, 5687 (2000).}

\bibitem{inouye99}
{S. Inouye, A.P. Chikkatur, D.M. Stamper-Kurn, J. Stenger, D.E. Pritchard,
and W. Ketterle, Science {\bf 285}, 571 (1999);
M.G. Moore and P. Meystre, Phys. Rev. Lett., {\bf 83}, 5202 (1999).}

\bibitem{piovella2001}
{N. Piovella, R. Bonifacio, B.W.J. McNeil, and G.R.M. Robb,
Opt. Commun. {\bf 187}, 165 (2001).}

\bibitem{thirunamachandran80}
{T. Thirunamachandran, {M}ol. Phys. {\bf 40}, 393 (1980);
D.P.~Craig and T.~Thirunamachandran, \textit{Molecular Quantum
Electrodynamics} (Academic Press, London, 1984).}

\bibitem{a+s}
{M.  Abramowitz and I. Stegun,
\textit{Handbook of Mathematical Functions}
(National Bureau of Standards, Washington, 1964).}


\bibitem{aneg}
{C.C.~Bradley, C.A.~Sackett, J.J.~Tollet, and 
R.G.~Hulet,   Phys. Rev. Lett. {\bf 75}, 1687 (1995); 
C.C.~Bradley, C.A.~Sackett, and R.G.~Hulet, Phys. Rev. Lett. 
{\bf 78}, 985 (1997);  C.A.~Sackett, H.T.C.~Stoof, and 
R.G.~Hulet, Phys. Rev. Lett. {\bf 80}, 2031 (1998).}

\bibitem{pitaevskii84}
{L.P. Pitaevskii, Zh. Eksp. Teor. Fiz. \textbf{39}, 423 (1984).}

\bibitem{sprsld} 
{M.~Kalinski, I.E.~Mazets, G.~Kurizki, 
B.A.~Malomed, K.~Vogel, and W.P.~Schleich, ``Dynamics of 
laser-induced supersolid formation in Bose-Einsten condensates'' 
cond-mat/0310480.}


\end{thebibliography}
\end{document}